\font\twelverm = cmr12 \font\tenrm = cmr10 \font\eightrm = cmr8
       \font\sevenrm = cmr7
\font\twelvei = cmmi12 \font\eighti = cmmi8
       \font\teni = cmmi10 \font\seveni = cmmi7
\font\twelveit = cmti12 
        \font\sevenit = cmti7
\font\twelvesy = cmsy10 scaled\magstep1
       \font\tensy = cmsy10 \font\eightsy = cmsy8 \font\sevensy = cmsy7
\font\twelvebf = cmbx12 \font\tenbf = cmbx10
        \font\sevenbf = cmbx7
\font\twelvecaps = cmcsc10 scaled\magstep1 \font\tencaps = cmcsc10
\font\twelvesl = cmsl12
\font\twelveit = cmti12
\font\twelvett = cmtt12
%
\textfont0 = \twelverm
       \scriptfont0 = \tenrm  \scriptscriptfont0 = \sevenrm
       \def\rm{\fam0 \twelverm}
\textfont1 = \twelvei
       \scriptfont1 = \teni  \scriptscriptfont1 = \seveni
       
\textfont2 = \twelvesy
       \scriptfont2 = \tensy \scriptscriptfont2 = \sevensy
       
\newfam\itfam \def\it{\fam\itfam \twelveit} \textfont\itfam=\twelveit
\newfam\slfam  \textfont\slfam=\twelvesl
\newfam\bffam  \textfont\bffam=\twelvebf
       \scriptfont\bffam=\tenbf \scriptscriptfont\bffam=\sevenbf
\newfam\ttfam  \textfont\ttfam=\twelvett
%
%
\rm
%
%
\hsize=6.5in
\vsize=9in
\raggedbottom
\baselineskip = 14 pt
\parskip=10pt plus 1pt
\rightskip=0pt  \spaceskip=0pt  \xspaceskip=0pt
\pretolerance=100  \tolerance=200
%
\overfullrule=0pt
%
               \def\makeheadline{\vbox to 0pt{\vskip-36.5pt
               \line{\vbox to8.5pt{}\the\headline}\vss}\nointerlineskip}
%
\dimen1=\baselineskip \divide\dimen1 by 2
\dimen2=\dimen1 \multiply\dimen1 by 2 \divide\dimen1 by 3
%
\nopagenumbers
\pageno=1
\headline={\ifnum\pageno=1 \hss\thinspace\hss
     \else\hss\folio\hss \fi}
%
\def\heading#1{\vfill\vbox to \dimen1 {\vfill}
     \centerline{\twelvecaps #1}
     \global\count11 = 0  
     \vskip \dimen1}
%
\count10 = 0
\def\section#1{\vbox to \dimen1 {\vfill}
    \global\advance\count10 by 1
    \centerline{{\number\count10}.\ \twelvecaps {#1}}
    \global\count11 = 0  
                  }
%
\def\subsection#1{\global\advance\count11 by 1
    \vskip \baselineskip
    \centerline{{\number\count10}.{\number\count11}.
    \ \it {#1}}
    \global\count12=0 
    \vskip -7pt}
%
\def\subsubsection#1{\global\advance\count12 by 1
     \vskip \baselineskip
     \centerline{{\number\count10}.{\number\count11}.{\number\count12}.
     \ \it {#1}}
     \vskip -7pt}
%
%
\def\refindent{\advance\leftskip by 24pt \parindent=-24pt}
%
\def\journal#1#2#3#4#5{{\refindent
                      {#1}        
                      {#2},       
                      {#3},       
                      {#4},       
                      {#5}.       
                      \par }}
%
\def\journetal#1#2#3#4#5{{\refindent
                        {#1}{ et al. }  
                        {#2},           
                        {#3},           
                        {#4},           
                        {#5}.           
                        \par }}
%

%
\def\inbook#1#2#3#4#5#6#7{{\refindent
                         {#1}         
                         {#2},        
                      in {#3},        
                     ed. {#4}         
                        ({#5}:        
                         {#6}),       
                       p.{#7}.        
                         \par }}
%

%


%
\def\privcom#1#2#3{{\refindent
                  {#1}        
                  {#2},       
                  {#3}.       
                  \par }}
%

%

%

%
\def\figcap#1#2{{\baselineskip=10pt \eightrm \narrower \noindent
                 \textfont0 = \eightrm \textfont1 = \eighti
                 \textfont2 = \eightsy
                 \scriptfont0 = \sevenrm  \scriptscriptfont0 = \sevenrm
                 \scriptfont1 = \seveni  \scriptscriptfont1 = \seveni
                 \scriptfont2 = \sevensy \scriptscriptfont2 = \sevensy
                 \vskip \baselineskip
                   Fig.\ {#1}.---  
                         {#2}      
                        \par}}
%

%
%
\def\etal{et al.}

%
\def\to {$\rightarrow$}
\def\smallneg {\kern-.08em}

\def\CO {{$^{12}$CO}}
\def\cO {$^{13}$\smallneg CO}
\def\cii {[C~{\tenrm II}]}
\def\ci {[C~{\tenrm I}]}
\def\oi {[O~{\tenrm I}]}
\def\cplus {C$^+$}
%
\def\h {$^{\rm h}$}
\def\m {$^{\rm m}$}
\def\s {$^{\rm s}$}
\def\ps {\rlap{\hbox{$^{\rm s}$}}{\hbox{.}}\kern0.15em}
\def\deg{$^{\circ}$}
\def\am{${'}$}
\def\as{${''}$}
\def\pas {\rlap{\hbox{$''$}}{\hbox{.}}\kern0.15em}
\def\pam {\rlap{\hbox{$'$}}{\hbox{.}}}
%
\def\cmarea {cm$^{-2}$}
\def\cmv {cm$^{-3}$}
\def\kms {km~s$^{-1}$}
\def\intens {ergs s$^{-1}$ cm$^{-2}$ sr$^{-1}$}
%
\def\nhtwo {$n_{{\rm H}_2}$}

\def\x{~$\times$~}
\def\by {~$\times$~}
\def\ee#1#2{${#1} \times 10^{#2}$}
\def\about {$\sim$~}
\def\lsim {$\rlap{\raise.4ex\hbox{$<$}}\lower.55ex\hbox{$\sim$}\,$}
\def\gsim {$\rlap{\raise.4ex\hbox{$>$}}\lower.55ex\hbox{$\sim$}\,$}
%

\def\far-ir{far-infrared}

\def\um {$\mu$m}
\def\dv {$\Delta V$}
\def\vlsr {$V_{LSR}$}

\def\Tastar {$T_A^*$}
\def\Tmb {$T_{MB}$}
\def\thetaonec {$\theta^1$C}

\def\orikl {Orion~KL}
\def\Go {$G_\circ$}

%
%
{\scriptfont0 = \sevenrm
\null\vskip -50.5pt
\centerline {University of Texas Astronomy Dept.\ Preprint No.\ 211 \hfill
astro-ph/9301009}
\vskip 36.5pt
\centerline {\twelvecaps EXTENDED CO(7\to6) EMISSION FROM WARM GAS IN ORION}
\vskip 12pt
\centerline {\tencaps
J. E. Howe,\footnote{$^ 1$}{\sevenrm \hskip -4pt Department of Astronomy,
University of Texas at Austin, Austin,
TX 78712}$^{,}$\footnote{$^2$}{\sevenrm \hskip -4pt Current address:
Department of Astronomy, University of Maryland, College Park, MD 20742}
D. T. Jaffe,$^1$
E. N. Grossman,$^{1,}$\footnote{$^3$}{\sevenrm \hskip -4pt
Current address: National Institute of Standards and Technology,
Div. 814.03, 325 Broadway, Boulder, CO 80303}
W. F. Wall,$^{1,}$\footnote{$^4$}{\sevenrm \hskip -4pt Current
address: NASA Goddard Space Flight Center, Code 685, Greenbelt, MD 20771}}
\vskip -3pt
\centerline {\tencaps
J. G. Mangum,$^1$
and G. J. Stacey\footnote{$^5$}{\sevenrm \hskip -4pt Department of Astronomy,
Cornell University, Ithaca, NY 14853}}
\vskip -2pt
\centerline{\sevenrm To appear in {\sevenit The Astrophysical Journal}}
\vskip -6pt
\centerline{\sevenit Received 1992 October 2; accepted 1992 December 1}
}
\vskip 6pt
\centerline {\tencaps ABSTRACT}
\vskip -8pt

{\baselineskip 12pt \narrower\narrower \scriptfont0 = \eightrm
\tenrm We mapped the quiescent emission from the 807 GHz J = 7\to6
transition of CO in Orion along a strip in R.A. extending from
0.7 pc west to 1.2 pc east of \thetaonec~Orionis.
The lines arise in warm gas with temperatures greater than 40~K.
The line brightness temperature is greater than 160~K in the
direction of \thetaonec, more than twice the dust temperature,
and still exceeds 35~K more than a parsec east of \thetaonec.
The lines are narrow, with a maximum velocity width of 7 \kms\
near \thetaonec\ and decreasing to 1.5--3 \kms\ at the map
boundaries.  The density of the emitting gas is greater than
$10^4$ \cmv\ and the column density exceeds $10^{21}$ cm$^{-2}$.
The correlation of the bright, narrow CO(7\to6) lines with 158 \um\
[C~{\eightrm II}] emission suggests that over the entire
region mapped, the narrow CO lines arise in warm
photodissociation regions excited by ultraviolet (UV) photons from the
Trapezium cluster.  Although the Trapezium stars lie in front of the
Orion~A molecular cloud, not all of the warm gas is at the cloud surface.
To the east of \thetaonec\ the CO(7\to6) lines
split into two velocity components (also seen in J = 2\to1 \cO\ emission)
which persist over several arcminutes.  Since only one of these
components can be on the surface, the other must arise
from a dense, UV-illuminated clump or filament within the molecular
cloud.  Comparison of the quiescent
CO(7\to6) emission to CO(7\to6) observed in a cross-map of the
energetic Orion~KL outflow shows that the luminosity of
shock-excited CO(7\to6) emission in Orion is
only a few percent of the luminosity of the widespread quiescent
CO(7\to6) emission.\par
\vskip -8pt
{
\noindent{\it Subject headings:\/} interstellar: matter---interstellar:
molecules---nebulae: Orion Nebula
\par}
}
\section {introduction}
Recent observations of galactic star-forming regions
indicate that ultraviolet (UV) photons from nearby or embedded
OB stars can penetrate dense molecular cloud cores over parsec
and greater scales, even though the interior regions appear well
shielded from external UV radiation by large average dust column
densities.  UV photons with energies greater than 11.3 eV
ionize atomic carbon, which radiates far-IR line
emission at 158 \um.  The bulk of the 158 \um\ \cii\ emission
arises in interface zones, or photodissociation regions (PDRs),
on the surfaces of dense molecular regions (\nhtwo\ \gsim\ $10^3$ \cmv)
 illuminated by the O or B stars
(Crawford \etal\ 1985; Stacey \etal\ 1985; Wolfire, Hollenbach,
\& \hbox{Tielens} 1989).  Maps of the far-IR 158 \um\ \cplus\ fine-structure
line in M17, W3, and NGC 1977 show that the \cii\ emission
extends more than a parsec into dense molecular gas (Harris
\etal\ 1987; Stutzki \etal\ 1988; Matsuhara \etal\ 1989; Howe
\etal\ 1991), although in uniform dense gas the dust should
attenuate the UV within a few hundredths of a parsec.  Modeling
of the intensity and distribution of the \cii\ emission in M17,
W3, and NGC 1977 shows that UV photons penetrate through a
relatively transparent interclump medium, illuminate the
surfaces of dense clumps or filaments, and produce the extended
\cii\ emission within the molecular clouds (Stutzki \etal\ 1988;
Howe \etal\ 1991) rather than only near the UV sources.

In a PDR, a warm molecular layer with temperatures of up to
several hundred degrees underlies the predominantly atomic
region where most of the \cii\ emission arises.  Emission
from rotational transitions of CO is an important
coolant of the molecular component of a PDR (Tielens \&
Hollenbach 1985a; Sternberg \& Dalgarno 1989) and dominates the
gas cooling when the density is \lsim\ $10^5$ \cmv\ (Sternberg
\& Dalgarno 1989).  Most of the CO line luminosity is emitted
by transitions from rotational states with
energies from 120 to 300 K (J = 6 to 10) above the ground
state (Burton, Hollenbach, \& Tielens 1990).  The CO(7\to6) line
is therefore a good tracer of the warm molecular component of
a PDR.  The small dipole moment of the CO molecule (0.11 Debye)
and its large abundance in molecular clouds ([CO]/[H$_2$]
\about$10^{-4}$) make the rotational lines easily excited and relatively
bright over a wide range of molecular gas temperatures and
densities and make it possible to use these lines to derive
the physical characteristics of the molecular gas.

The extensive distribution of PDRs in the M17, W3, and
NGC 1977 molecular clouds suggests that widespread
CO(7\to6) emission, like \cii, may be common in high-mass
star-forming regions.  The Orion A region
is a natural choice for a study of the extended spatial
distribution of high-excitation molecular gas in a
star-forming cloud.  It is relatively nearby (distance
\about470~pc; \hbox{Genzel} \etal\ 1981), giving a linear scale of 0.14
pc\ arcmin$^{-1}$, and bright submillimeter CO lines have been
observed from quiescent gas and from the energetic
compact outflow at Orion KL.  Nearly all of the previous
observations of the CO J = 7\to6 and J = 6\to5 emission lines
in Orion have been confined to the inner few arcminutes centered
on the Orion~KL nebula (Fetterman \etal\ 1981; Goldsmith \etal\ 1981;
Buhl \etal\ 1982; Koepf \etal\ 1982; Schultz \etal\ 1985;
Graf \etal\ 1990).  Schmid-Burgk \etal\ (1989) mapped
the CO(7\to6) emission from a 6\am\by8\am\ region encompassing
Orion~KL and the Trapezium OB cluster, but the observations were
spatially chopped with an amplitude of only \about2\am, not much
larger than their 98\as\ beamsize, and therefore required
corrections for substantial contamination from emission in the
reference beam.  The extended CO(7\to6) map of Schmid-Burgk \etal\
was derived from the data by a bootstrapping method whereby the
reference beam of successive observations is placed at the
position of the previous observation.  While the
bootstrapping technique provides adequate information about the
compact bright core of the Orion region, it is insensitive
to low-level extended emission.  In the most extreme
case, any emission more extended than the mapping limits is
completely subtracted from the reconstructed map.

We present here the results of a search for extended
CO(7\to6) line emission in the Orion A molecular cloud using the
University of Texas submillimeter laser heterodyne receiver.  We
mapped the large-scale distribution of quiescent CO(7\to6) emission
along a cut across the Trapezium cluster extending more than 13\am\
(nearly 2 pc) in R.A. \ In contrast to the practice more common in
high-frequency submillimeter heterodyne spectroscopy of spatially
chopping with 2\am~--~6\am\ throws, we employed the position-switching
method where the telescope is slewed to a reference position well away
from the source position to insure against contamination from
emission at the reference position.  Our sensitivity to extended
low-level line emission is limited, therefore, only by the
receiver noise and the opacity of the atmosphere.  We describe in
\S~2 the receiver, observing method, and calibration techniques.
The CO(7\to6) spectra and maps are presented in \S~3.  The
physical parameters and energetics of the molecular gas are
discussed in \S~4. In \S~5 we summarize the results of the
investigation and compare our findings with observations of
other high-luminosity star-forming regions.

\section {observations}
We observed the J=7\to6 rotational transition of CO
($\nu$  = 806.6517 GHz, $\lambda$ = 371.6504 \um) using the University
of Texas submillimeter laser heterodyne receiver mounted at the Cassegrain
focus of the 10.4 m telescope of the
Caltech Submillimeter Observatory\footnote{$^{\tenrm 6}$}{{\eightrm The CSO
is operated by
the California Institute of Technology under funding from the
National Science Foundation, contract AST-9015755.}}
(CSO) on 1990 Dec 2.  The receiver employs a
liquid-nitrogen cooled Schottky diode mounted in
an open-structure 90\deg\ corner-reflecting mixer
block (e.g., Kr\"autle, Sauter, \& Schultz 1977).
The local oscillator is an optically pumped
submillimeter molecular laser operated at 802.986 GHz, which
gives a first intermediate frequency of 3.67 GHz.  This signal
is further mixed down to 1 GHz before sampling by the CSO
backend 500 MHz acousto-optical spectrometer (AOS).

The receiver noise temperature measured on the telescope
was 3850~K (double sideband) and the measured zenith atmospheric
transmission at 807 GHz was 0.32 (after correcting for
the unequal atmospheric transmission at the signal and image
sideband frequencies).  The total scatter in the zenith
transmission measured during the observations was less than
$\pm$~0.02.  The absolute intensity scale was calibrated by observing
blackbody loads at ambient and liquid-nitrogen temperatures
placed in the signal beam path.  We calibrated at each position
observed, or every five minutes when we integrated at a single
position for longer durations.  The atmospheric transmission
along the line of sight to the source was measured during each
calibration series from the broadband signal received from the
sky.  The observations were taken in position-switching mode
with the reference position 30\am\ west of the mapping center
position (\thetaonec\ Orionis or \orikl).  Only first-order baseline
corrections were applied to the spectra presented here, since both
the atmosphere and the receiver were stable during the observations.
The absolute pointing was
established by maximizing the continuum signal received from a
planet (Mars or Jupiter).  We estimate that the absolute
pointing accuracy is $\pm$~7\as, with the relative pointing in the
mapping observations accurate to $\pm$~2\as\ or less.

The telescope beam efficiency was measured using the
continuum signals from Jupiter (diameter $D = \sqrt{41'' \times 39''}
 = 40''$) and Mars ($D = 18''$), which was near opposition.  Using
a 372~\um\ brightness temperature of 143~K for Jupiter [Hildebrand
\etal\ (1985), corrected for Jovian atmospheric absorption bands
of HCN and PH$_3$ in the signal and image sidebands (Lellouch,
Encrenaz, \& Combes 1984)] and 220~K for Mars at opposition
(Wright 1976; Wright \& Odenwald 1980), we derive a coupling
efficiency to Jupiter $\eta_{Jup} = 0.24$, and a coupling efficiency to
Mars $\eta_{Mars} = 0.09$.  The beam shape was determined from Gaussian
profiles fitted to cross-maps of Jupiter and Mars.  The mapping
data for both planets are well modeled by a beam that is the
superposition of two Gaussian beams:  a main beam of full width
at half maximum amplitude FWHM = 20\as\ and a broader error beam of FWHM
= 90\as, with an amplitude ratio of the main beam to the
error beam of 9.4 to 1.  These beam
parameters also
simultaneously reproduce the measured efficiencies on Jupiter
and Mars, and give a total beam efficiency (the efficiency
measured on a source that uniformly fills the sum of the two
beam components) $\eta_{tot} = 0.61$, which compares well with
coupling efficiencies to the Moon measured on earlier observing runs,
$\eta_{Moon} = 0.65$.
The coupling efficiency of this beam to Gaussian sources with FWHM =
30\as, 60\as, and 90\as\ is 0.18, 0.30, and 0.40, respectively.  The
effective solid angle subtended by the beam is $\Omega_{beam}$ =
\ee{3.04}{-8}~sr, equivalent to the solid angle of a single 34\as\ FWHM
Gaussian beam.

We also observed the J = 2\to1 transition of \cO\ ($\nu$ =
220.3987 GHz, $\lambda$ = 1.3602 mm) with the CSO facility 1.3 mm SIS
receiver, also in position-switching mode, with the reference
position 36\am\ north of the source (a position previously determined
to be free of low-J CO emission).  We used the standard
chopper-wheel calibration method (Penzias \& Burrus 1973) to
correct for atmospheric attenuation and to set the thermal
scale.  The receiver noise temperature was \about420~K (single
sideband) and the atmospheric transmission was 0.92 or higher.
The beam at 220 GHz is a single Gaussian with FWHM = 33\as\
($\Omega_{beam}$ = \ee{2.86}{-8}~sr) and a main beam efficiency
$\eta_{MB} = 0.72$.

For the submillimeter observations we used the 1024-channel
500 MHz AOS, which gave a velocity resolution of 0.37 \kms\
and a total velocity coverage of 186 \kms.  For the millimeter
observations we used the 1024-channel 50 MHz AOS and the 500
MHz AOS, which gave velocity resolutions of 0.13 \kms\ and 1.34
\kms\ and total velocity coverages of 68 \kms\ and 680 \kms,
respectively.  We checked the frequency calibration and spectral
resolution of the backend AOS's throughout the millimeter and
submillimeter observations.

\section {results}
We observed bright, narrow CO(7\to6) lines (velocity width \dv\
\about1.5~--~7.0 \kms) along a Right Ascension cut at the
Declination of \thetaonec~Ori [R.A.(1950) = 05\h\ 32\m\ 49\s, Decl.(1950)
= $-$05\deg\ 25\am\ 16\as] extending from 300\as\ west to 525\as\ east of
\thetaonec.  The CO(7\to6) spectra are shown in Figure 1.  We also show
in Figure 1 \cO(2\to1) spectra observed at the same positions for
comparison.  The horizontal scale of each spectrum extends from
\vlsr\ = $-$5 to $+$25 \kms.  The vertical scale gives
the Rayleigh-Jeans antenna temperature of the CO(7\to6) emission
corrected for telescope losses and atmospheric attenuation
[\Tastar\ in the notation of Kutner \& Ulich (1981)] divided by a
807 GHz beam coupling efficiency factor of 0.4.
The beam coupling factor is appropriate for
sources such as Orion with brightness distributions which vary on
moderately extended (\about90\as) scales and is a good
compromise between the coupling factors of very large uniform
sources ($\eta_{tot}$ \about0.6) and sources smaller than 60\as\ ($\eta$
\about0.3).  The \cO(2\to1) spectra are plotted in units of Rayleigh-Jeans
main beam brightness temperature \Tmb, multiplied by a factor of 3
to facilitate comparison with the J = 7\to6 spectra.
The spacing between the observed positions is 75\as, except
that time constraints prevented us from observing the CO(7\to6)
spectrum at the positions 75\as\ east
and 75\as\ west of \thetaonec.  The CO(7\to6) intensity map of
Schmid-Burgk \etal\ (1989) indicates that the position of maximum
peak line temperature, which we estimate is at least a factor of
1.15 higher than the line temperature toward \thetaonec, is
\about40\as\ west of \thetaonec.  Our own \cO(2\to1) observations also
suggest that the CO(7\to6) line temperature peaks at that position.

\vfill\eject

\null\vskip 195.0pt
\figcap{1}{CO(7\to6) (upper spectra) and \cO(2\to1) (lower spectra)
emission lines observed along a cut in R.A. in Orion.
The 0\as\ position is at \thetaonec~Ori [R.A.(1950) = 05\h\ 32\m\ 49\s,
Decl.(1950) = $-$05\deg\ 25\am\ 16\as].  The spacing between adjacent boxes
is 75\as, and the R.A. offset from \thetaonec\ in arcseconds is given at
the top of each box.  The horizontal scale of each spectrum extends
from \vlsr\ = $-$5 to $+$25 \kms.  The vertical scale gives the
Rayleigh-Jeans antenna temperature \Tastar\ of the CO(7\to6) spectra
divided by a 807 GHz beam coupling factor of 0.4.  The \cO(2\to1)
spectra are plotted in units of 3\x\Tmb.}
\vskip 6pt
The CO(7\to6) lines are brightest within a few arcminutes of
the Trapezium, with intensities ranging from 60 to 140~K and
requiring molecular gas temperatures of at least 160~K near the
exciting stars, substantially higher than the 55~--~85~K color
temperature of the far-IR dust emission from the same region
(Werner \etal\ 1976; Jaffe \etal\ 1984).
To the west of \thetaonec\ the line temperature
drops from \about40~K at 225\as\ west to \about10~K at 300\as\ west,
implying minimum kinetic temperatures of 57~K and 25~K respectively.
The CO(7\to6) line intensity is \about140~K at \thetaonec\ and
exceeds 50~K 300\as\ (0.7 pc) to the east.  Emission is still detected
525\as\ (1.2 pc) east and 300\as\ west of \thetaonec, with line temperatures
of \about25~K and \about10~K, respectively, and most likely extends
farther in both directions.  The line widths also decrease
with distance from \thetaonec.  At \thetaonec\ the linewidth
is \about7 \kms,
while 150\as\ and 225\as\ to the east the linewidth is \about5 \kms.
At 300\as\ east of \thetaonec\ the linewidth has narrowed
to only 1.8 \kms.

The velocity structure of both the CO(7\to6) and the
\cO(2\to1) spectra becomes increasingly complex east of \thetaonec,
exhibiting multiple velocity components whose brightnesses vary
from position to position.  At positions beyond 375\as\ east of
\thetaonec, the CO(7\to6) line profiles break up into two velocity
components with narrow velocity widths (\dv\ \about1.5~--~2.0 \kms)
and line temperatures of 25~--~30~K
(implying minimum gas temperatures of 41~--~47~K).  Figure 2
compares the \cO(2\to1) and CO(7\to6) line profiles for a spectrum
synthesized by coadding spectra at 450\as\ east and 525\as\ east of
\thetaonec\ to increase the signal to
noise ratio of the CO(7\to6)
spectrum.  The CO(7\to6) profile has velocity components at 6
and 9 \kms\ with line temperatures \about25~K.  The \cO(2\to1) spectrum
also has components at 6 and 9 \kms, but with line
temperatures differing by factor of \about3 (\Tmb\ = 12~K and
\hbox to 6.5in{4~K respectively).  In addition, there is a bright
(\Tmb\ = 10~K) \cO(2\to1) component}\vadjust{\eject}

\null\vskip 199.3pt
\figcap{2}{CO(7\to6) and \cO(2\to1) synthesized spectra
generated by the coaddition of positions 450\as\ and 525\as\ east of
\thetaonec~Ori.  The vertical temperature scale gives \Tastar\ divided
by 0.4 for the CO(7\to6) spectrum and 2\x\Tmb\ for the \cO(2\to1)
spectrum.  The dotted lines mark the center velocity of the 6 and
9 \kms\ \cO\ line components.  The 1-$\sigma$ noise per channel for
the CO(7\to6) spectrum is also indicated.}
\vskip 6pt
\noindent
at 11 \kms\ with no observed counterpart
in CO(7\to6).  More complete observations of the \cO\ emission
sampled every 15\as\ in R.A. from 300\as\ east to 885\as\ east of \thetaonec\
show that the velocity components persist on spatial scales of
\about60\as~--~120\as.

In the following section we discuss the physical conditions
and morphology of the emitting gas implied by the observations,
and the excitation of the submillimeter CO lines.

\section {discussion}
\subsection {Emission from Quiescent Gas}
\subsubsection {Gas Excitation Near Ori \thetaonec }
The local CO(7\to6) intensity maximum near \thetaonec\
indicates that the UV photons from the Trapezium stars are the
excitation source for the CO(7\to6) emission.  The intense flux of
Lyman continuum photons, contributed primarily by the O6 star \thetaonec,
dissociates and ionizes gas at the surface of the background
Orion A cloud, creating an H~{\tenrm II} region \about5\am\ in diameter
(Zuckerman 1973; Balick, Gammon, \& Hjellming 1974).  Between the
H~{\tenrm II} region and the bulk of the molecular cloud lies a
photodissociation region of partially ionized atomic and neutral
molecular gas where the energetics are dominated by UV photons
with energies less than the Lyman limit.  Behind \thetaonec, where the
local UV flux $G$ is roughly $10^5$ \Go\ at the cloud surface (\Go\ =
\ee{1.6}{-3}\ ergs s$^{-1}$ \cmarea, the average interstellar flux of
UV photons with energies of 6~--~13.6 eV), a PDR model
well explains the observed far-IR \oi, \ci,
and \cii\ intensities (Tielens \& Hollenbach 1985a,b).
The observational result that the brightness temperature
of the CO(7\to6) line near \thetaonec\ is more than twice the dust
\hbox to 6.5in{temperature in the same region, a condition only observed in
molecular clouds with nearby}\vadjust{\eject}

\null\vskip 176.6pt
\figcap{3}{The LVG model for a \CO\ column density per unit velocity
interval $N_{\rm CO} / {\rm d}V$ = $10^{18}$ \cmarea\ km$^{-1}$~s, plotted
in the temperature-density plane. Solid contours give the Rayleigh-Jeans
brightness temperature of the J = 7\to6 transition of CO while the dashed
contours denote the brightness temperature ratio of CO(7\to6) to
\cO(2\to1).  The gray regions 1, 2, 3, and 4 show the intersection of
the value of
CO(7\to6) \Tastar\ $/0.4$ ($\pm 1\sigma$) with the CO(7\to6) \Tastar\ $/0.4$
to \cO(2\to1) \Tmb\ ratio ($\pm 1\sigma$) for positions (525,0),
(150,0), ($-$150,0), and (0,0), respectively.}
\vskip 6pt
\noindent
UV sources, clearly rules out
collisions with warm dust as the heating mechanism for the warm
CO.  In a PDR, however, the high flux of UV photons can directly
heat the gas to temperatures higher than the dust component
through collisions with electrons photoejected from dust grains
and by collisional de-excitation of vibrationally excited,
UV-pumped H$_2$.  The similar morphology of the CO(7\to6) emission
(Schmid-Burgk \etal\ 1989) and the 158 \um\ \cii\ emission
(Stacey \etal\ 1992) in Orion strongly suggests a common origin
for the excitation of these lines (Genzel \etal\ 1989; Stacey
\etal\ 1992).  The narrow width of the CO(7\to6) lines also makes it
unlikely that shocks are important in heating the gas near the
Trapezium stars.

\subsubsection {Temperature, Density, and Column Density Estimates}
We can use the observed CO(7\to6) and \cO(2\to1) line
intensities and intensity ratios to estimate the temperature,
density , and column density of the molecular gas. For gas
densities of more than a few $10^3$ \cmv\ and temperatures higher
than \about40~K, the intensity of the CO(7\to6) line emission depends
more sensitively on the physical conditions in the emitting gas
than do the millimeter CO transitions (the J = 7 rotational
level lies 155~K above the ground state and the critical
density of the J = 7\to6 transition is $n_{crit}$ \about\ee{6}{5}\ \cmv).
Although there is ample evidence that the molecular gas in Orion is far
from uniform, we assume for the moment that the CO and \cO\ lines are
emitted from a uniform gas at a single temperature and density.
The uniform model results set lower limits to the actual
temperature, density, and column density of the PDR layers, as
we discuss later.  We adopt a value of 60 for the [\CO]/[\cO]
abundance ratio in Orion [see Blake \etal\ (1987) for a
discussion of abundances in Orion].  We compare the observed
line intensities and ratios with line intensities computed from
a non-LTE large velocity gradient (LVG) radiative transfer
model (Scoville \& Solomon 1974; Goldreich \& Kwan 1974) to
estimate the range of the physical conditions of the
emitting gas.  An LVG model is the simplest way to include
non-LTE effects and local radiative trapping in the estimation of
the physical conditions from the line intensities.  The model
employs the H$_2$--CO collision rate constants of Schinke \etal\ (1985)
and Flower \& Launay (1985) and assumes a plane-parallel cloud geometry.
Figure 3 shows the LVG model result for a \CO\ column density per unit
velocity interval $N_{\rm CO} / {\rm d}V$ = $10^{18}$ \cmarea\ km$^{-1}$~s.
The model results indicate
that uniform gas with temperatures of 40~--~200~K, molecular
hydrogen densities \nhtwo\ of $10^3$~--~$10^5$ \cmv, and CO
column densities $N_{\rm CO}$ of \ee{2}{17}~--~\ee{4}{18}\ \cmarea\ can
reproduce the entire observed range of detected CO(7\to6) line intensities
(25~K $<$ \Tastar\ /(0.4) $<$ 140~K) and CO(7\to6) to \cO(2\to1) intensity
ratios (which range from 2 to 15).  For example, at the position
150\as\ west of \thetaonec, $N_{\rm CO}$ \about\ee{2}{18}\ \cmarea,
\nhtwo\ $>$ \ee{6}{4}\ \cmv, and
$T_{gas}$ \about110~K produces the observed line intensity of
98~K and CO(7\to6) to \cO(2\to1) ratio, while at the position 450\as\
east of \thetaonec, the best LVG model fit for the 9 \kms\ velocity
component is obtained with $N_{\rm CO}$ \about\ee{2}{17}\ \cmarea\ and
\nhtwo\ varying from $10^3$ to $10^{4.5}$ \cmv\ as $T_{gas}$
goes from 200 to 50~K.  At a few positions, any
density \nhtwo\ $>$ $10^4$ \cmv\ gives
an acceptable line intensity for one combination of temperature and
column density within the range of temperatures and column densities
given above.  For the bright \cO(2\to1) velocity components (\Tmb\
$>$ 10~K) which have no observable counterpart in CO(7\to6), such as
the 11 \kms\ component at positions farther than 150\as\ east of
\thetaonec, CO column densities of about $10^{18}$ \cmarea\ and low
temperatures (less than 40~K) are required for excitation
conditions that produce the observed \cO\ line temperatures without
simultaneously increasing the CO(7\to6) line temperature to an
observable level.

\subsubsection {Excitation of Extended Emission}
The large-scale distribution of \cii\ emission, especially
east of \thetaonec, is similar to the distribution of
CO(7\to6) emission (Fig.\ 4) and suggests that the
\cii\ and CO lines share a common PDR origin across the {\it entire\/}
region mapped and not only in the immediate vicinity of the Trapezium.
Mapping of the 158 \um\ \cii\ line across the Trapezium
region shows that the \cii\ line arises from a region more than
30\am\ (4 pc) in extent (Stacey \etal\ 1992), and indicates the
presence of a large UV flux (higher than 100 \Go) beyond the limits
of our CO(7\to6) observations.  A more detailed fully-sampled map of
the \cii\ emission in the inner few arcminutes around \thetaonec\ shows
that the \cii\ emission peaks just west of the Trapezium
(Stacey \etal\ 1992), as does the CO(7\to6) emission.  The steep
decline in \cii\ emission to the west of \thetaonec, in the same region
where the CO(7\to6) emission is at a maximum, is a consequence of
the high density of the gas (\nhtwo\ $>$ $10^5$ \cmv) west
of the Trapezium traced by clumps of CS(2\to1) emission (Mundy
\etal\ 1986, 1988). At densities \gsim\ $10^5$ \cmv, the
self-shielding of the CO from dissociating UV radiation raises
the column density of warm
CO in the PDR relative to the column
density of \cplus\ and increases the relative strength of
the CO emission to the \cii\ emission (Wolfire \etal\ 1989).

The CO(7\to6) emission in the region \about1 pc east of \thetaonec\
probably also arises in PDRs, since the \cii\ emission from that
region indicates the presence of a PDR there (Stacey \etal\ 1992).
The flux of UV photons from the Trapezium stars in this region
is at least 100~\Go, based on the observed \cii\ line intensity,
and could be as much as a few $10^3$~\Go.  The UV flux
1 pc east of \thetaonec\ is similar to the UV flux in the
reflection nebula NGC 2023, where
\hbox to 6.5in{bright CO(7\to6) lines
(\Tmb\ \about70~K) are emitted from a PDR (Jaffe \etal\ 1990).
Thus, a}\vadjust{\eject}

\null\vskip 176.0pt
\figcap{4}{Comparison of the integrated intensity over all velocity
components of the CO(7\to6), \cO(2\to1), and 158 \um\ [C~{\sevenrm II}]
emission along the \thetaonec~Ori cut.  The peak intensities are
\ee{6.1}{-4}, \ee{1.8}{-6}, and \ee{4.2}{-3}, respectively, in units of
\intens.  The [C~{\sevenrm II}] data are from Stacey \etal\ (1992).
The peak intensity of the CO(7\to6) emission along
the R.A. cut is estimated from an extrapolation based on the map
of Schmid-Burgk \etal\ (1989).  The CO(7\to6) peak lies between
positions 0\as\ and $-$150\as.}
\vskip 6pt
\noindent
PDR origin for the quiescent (\dv\ $<$ 2 \kms)
extended CO(7\to6) emission in Orion, even in regions of moderate
UV photon flux (100 to 1000 \Go), is not without precedent.

\subsubsection {Evidence for Emission from Clump Surfaces}
The coexistence of the 6 and 9 \kms\ CO(7\to6) velocity
components over arcminute length scales makes it unlikely that both
components are at the surface of the molecular cloud; more
likely the components are emitted by spatially distinct clumps
or filaments separated along the line sight and illuminated by
the UV flux from the Trapezium. Observations and models of M17
(Harris \etal\ 1987; Stutzki \etal\ 1988), and W3 and NGC 1977
(Howe \etal\ 1991), demonstrate that PDR emission from the
surfaces of dense clumps or filaments can extend over parsec
length scales through molecular clouds.  Furthermore, Tauber \&
Goldsmith (1990) find that CO and \cO\ J = 3\to2 intensities and
line profiles in Orion are poorly fit by models with uniform
gas, but are better explained by a model composed of spherical
clumps with radial temperature and density gradients illuminated
by an external UV field.  On size scales of a
parsec or more, the \cO(1\to0) map by Bally \etal\ (1987)
shows that the Orion~A cloud is composed of a
highly nonuniform complex of filaments and clumps.
Ziurys \etal\ (1981) and Batrla \etal\ (1983) observed in the
(1,1) and (2,2) inversion lines of NH$_3$ a series of clumps in the
Orion ridge with \about0.1 pc diameters and smaller.  At still
smaller scales, interferometric maps of the emission from the
J = 2\to1 transition of CS at 7\pas5 resolution reveal four
distinct clumps \about30\as\ west of the Trapezium with diameters
ranging from 15\as to 50\as\ (0.04~--~0.12 pc) and densities greater
than $10^5$ \cmv\ (Mundy \etal\ 1986, 1988).  Our own \cO(2\to1) and
CO(7\to6) observations reveal multiple velocity components
persisting over \about1\am\ length scales and with varying excitation
conditions, particularly in the region a parsec or more east of \thetaonec.
Observations of the C109$\alpha$ carbon recombination line in Orion~A
at a spatial resolution of 2\pam6 show two velocity
components (\vlsr\ \about6 and 10~--~11 \kms) at positions 170\as\ east
and 327\as\ east of \thetaonec\ (Jaffe \& Pankonin 1978).  A spectrum of the
158 \um\ \cii\ line 4\am\ east of \thetaonec\ at 0.8 \kms\ spectral
resolution also shows two components, at \vlsr\ \about6 and 11 \kms\
(Boreiko, Betz, \& Zmuidzinas 1988).

The temperature, density, and column density
of the molecular gas must be higher in a nonuniform,
clumpy medium than the results of the uniform, single
temperature and density models suggest.  In a clumped gas, the
beam area filling factor can decrease from its value for a smoothly
varying gas.  The observed peak brightness temperatures of the
CO(7\to6) lines could then underestimate the true excitation temperature
of the gas and imply kinetic temperatures higher than the 40~K
brightness temperatures of the region \about1~pc east of \thetaonec.  The
local CO column densities within the clumps must also increase
to preserve the average value sampled within a beam diameter.
The higher gas temperatures implied by a clumped gas decrease
the fractional abundance of \cO\ molecules in the J = 2 state by
populating higher energy states, increasing further the local CO
column density required by the observed \cO(2\to1) line
intensities. The density of the gas in the clumps must then
increase so that the larger CO column densities do not require
excessively large clump diameters.  In a clumped or filamentary
gas, then, the CO(7\to6) and \cO(2\to1) observations require the
results of the simple LVG models to be interpreted strictly as
lower limits to the actual temperatures, densities, and column
densities of the emitting regions.  Indeed, the observed
integrated intensities of the extended component of the CO(7\to6)
emission lines imply densities \gsim\ $10^5$ \cmv\ if the
lines arise in PDRs (Burton, Hollenbach, \& Tielens 1990).  Even
if the CO is heated by collisions with warm dust, a gas density
greater than $10^5$ \cmv\ is required for the collisional
heating rate to equal the luminosity of
the CO emission.

Although the evidence supports the contention that the
CO(7\to6) lines a parsec east of \thetaonec\ arise in PDRs on the
surface of embedded clumps or filaments, a definitive answer as to
the excitation mechanism requires further observations.  The color
temperature of the far-IR dust continuum emission from that
region, as derived from IRAS data, ranges from \about35~K to 65~K,
depending on the power of the dust emissivity ($1 \leq \beta \leq 2$,
where we assume the dust emissivity varies with
wavelength as $\lambda^{-\beta}$).
Since the brightness of the CO(7\to6) lines in the same region
requires gas temperatures of at least 40~K, we can only say that
$T_{gas} \approx T_{dust}$, leaving open the possibility that the gas
is heated by collisions with the dust.
Observations such as the high resolution maps of CS emission
(Mundy \etal\ 1986, 1988), however, support a clumpy morphology to the
gas, in which case the gas temperature could be considerably
higher than the lower limit derived here.  If the gas is indeed
clumped on scales smaller than
the 807 GHz beam, observations at
higher spatial resolution should detect CO(7\to6) line
temperatures higher than the 25~K brightness temperatures we
observe. Observations of the 158 \um\ \cii\ line in this region
at a spectral resolution sufficient to resolve the velocity
components seen in the CO(7\to6) and \cO(2\to1) lines could
determine if the velocity and width of the \cii\ line emission
are the same as one, both, or neither of the CO(7\to6) velocity
components and would indicate a PDR origin of the emission if
the CO(7\to6) and \cii\ linewidths and velocities were the same.
The lack of a CO(7\to6) line component at the $+$11 \kms\ velocity
of one of the C109$\alpha$ and \cii\ line components may indicate a low
density ($n$ \about$10^3$ \cmv) for the gas emitting those lines, and
therefore a very subthermal population of CO in the J = 7 state.
Observations of spectrally resolved 158 \um\ \cii\ lines at the
velocities of the CO(7\to6) line components and observations at
higher spatial
\hbox to 6.5in{resolution of CO(7\to6) brightness temperatures
higher than the dust temperature in the}\vadjust{\eject}

\null\vskip 155.5pt
\figcap{5}{Cross-map of CO(7\to6) spectra observed towards
Orion~KL.  The (0,0) position is at R.A.(1950) = 05\h\ 32\m\
47\ps0 and Decl.(1950) = $-$05\deg\ 24\am\ 25\as, and the spacing
between the observed positions is 10\as\ in R.A. and in Decl.  The
horizontal scale of each spectrum extends from \vlsr\ = $-$80 to
$+$100 \kms.  The vertical scale of each spectrum extends from
$-$100 to 500~K in units of \Tastar\ divided by a beam coupling
factor of 0.2, which is appropriate for the wing emission but not
for the line core.  The R.A. and Decl.\ offsets in arcseconds from
the (0,0) position are indicated in the upper left corner of
each spectrum.}
\vskip 6pt
\noindent
same region would
conclusively demonstrate the PDR origin of the emission, rather
than heating by shocks or by collisions with warm dust.

\subsection {Emission from Shocked Gas:  Orion~KL}
About 1\am\ north of \thetaonec\ lies the energetic outflow source,
Orion~KL.  In marked contrast to the narrow velocity width and
wide spatial extent of the quiescent PDR emission excited by the
Trapezium cluster, the Orion~KL region exhibits extremely
energetic mass motions [\dv\ \about100 \kms\ full width to zero
power (FWZP) in the CO(7\to6) emission lines] confined to a
region less than 1\am\ in extent.  Figure 5 shows the
CO(7\to6) spectra observed towards Orion~KL.  The cross-map is
centered at R.A.(1950) = 05\h\ 32\m\ 47\ps0 and Decl.(1950) =
$-$05\deg\ 24\am\ 25\as, and the spacing between the observed positions is
10\as\ in R.A. and in Decl.  The horizontal scale of each spectrum
extends from \vlsr\ = $-$80 to $+$100 \kms.  Linear baselines were
removed from the spectra, which have been smoothed to a spectral
resolution of \about1.5 \kms\ per channel.  The vertical scale of
each spectrum extends from $-$100 to $+$500~K in units of \Tastar\
divided by an efficiency factor of 0.20, which is appropriate
for the coupling of the beam to a Gaussian source with FWHM
\about35\as.  This size scale is relevant only to the wing emission
and not to the line core, for which the temperature scale produces
artificially high temperatures (by approximately a factor of 2).

The most striking aspect of the wing emission is its absolute
brightness.  At all positions within
10\as\ of the map center, the line temperature exceeds 200~K even
for emission 20 \kms\ blueward of the line center velocity at 9
\kms, implying a kinetic temperature of at least 220~K.  The
line temperature of similarly redshifted emission at the same
positions is about 100~K less than the blueshifted emission.
Wing emission at velocities more than $\pm$ 20 \kms\ from the line
center velocity traces the high velocity flow seen in the millimeter CO
transitions, and is not likely affected by the hot low-velocity
flow seen in the 3~mm rotational transitions of SiO and SO [\dv\
(FWZP) \about35 \kms; Genzel \& Stutzki 1989].  At 20\as\ south of
KL, the wing emission has decreased enough so that the bright
spike feature characteristic of extended ridge emission becomes
readily apparent. Significant wing emission is still present 20\as\
north of KL, in agreement with the five-point CO(7\to6) cross-map
of KL obtained at 45\as\ spacing by Schmid-Burgk \etal\ (1989).
The integrated intensity of the blue wing emission generally
exceeds that of the red wing, a feature also seen in the
2.1 \um\ H$_2$ v = 1\to0 S(1) line emission from hot shock-excited
gas in the KL region (Nadeau \& Geballe 1979), indicating
that the excess blue emission in the H$_2$ line profiles may not be
an artifact of higher extinction of the red wing emission.
Excess blue wing emission is also seen in the line profiles of
shock-excited 63 \um\ \oi\ emission from the KL region (Werner
\etal\ 1984).  We note that although the line core emission appears
to be weakest at the central position in the cross-map, further
observations are needed to confirm this result.

\subsubsection {Outflow Size and Morphology}
The wing emission is spatially compact and
exhibits marked differences in both the absolute and relative
intensities of the red- and blue-shifted wing emission at
adjacent positions in the map.  Although the
high-velocity emission extends beyond the limits of the region
mapped, particularly to the north and west, the extent of the
emission to the half-power level is \about40\as\ in R.A. and \about35\as\
in Decl. This extent implies a source size less than 35\as\ in R.A. by 28\as\
in Decl.\ if the emission has a Gaussian distribution, consistent
with 35\as\ resolution CO(6\to5) mapping results (FWHM
\about45\as\by35\as; Koepf \etal\ 1982).  For a source of this size,
roughly 75 percent of the received power is detected by the 20\as\
main beam.  The CO(7\to6) emission region lies within the \about75\as\
FWHM region emitting 2 \um\ H$_2$ rovibrational lines (Beckwith
\etal\ 1978) and far-IR CO lines (Watson \etal\ 1985).  Our data
do not unambiguously indicate a bipolar structure to the outflow,
so the separation of the
centroids of the redshifted and the blueshifted emission
distributions must be small.  The morphology of the high-velocity
flow traced by the CO(7\to6) emission is consistent with
the data of Erickson \etal\ (1982), which show that the extent
of the redshifted and blueshifted CO(3\to2) emission contours is
\about30\as~--~40\as\ to the half-power level, that the centroids of the red
and blue lobes of the outflowing gas are separated by \about12\as, and
that the integrated intensity of the blue wing is greater than
the integrated intensity of the red wing.

\subsubsection {Density and Column Density Estimates}
The CO(7\to6) wing emission is probably emitted by cooling
postshock gas downstream of the hot ( \about2000~K) gas traced by
2~\um\ H$_2$ thermal rovibrational line emission (Beckwith \etal\ 1978;
Beckwith, Persson, \& Neugebauer 1979; Scoville \etal\ 1982;
Beckwith \etal\ 1983).  The bulk of the column density of the
gas emitting the CO(7\to6) line must be colder than the \about750~K
temperature characteristic of the gas emitting the far-IR CO
lines detected by Watson \etal\ (1985), since the integrated
intensity of the CO(7\to6) wing emission is about an order of
magnitude greater than can be emitted by the hotter component;
molecular gas at a temperature of 750~K or higher with a
sufficient column density to emit the observed CO(7\to6) line
intensity would also emit far-IR CO line emission much brighter
than observed.  The column density of the gas emitting the bulk
of the CO(7\to6) wing emission is at least \ee{5}{22}\ \cmarea\ [where we
use the non-LTE CO emission coefficients of McKee \etal\ (1982) and
assume an excitation temperature of \about250~--~500~K, density
\nhtwo\ \about$10^5$~--~$10^6$ \cmv, and CO abundance of $10^{-4}$
relative to H$_2$], substantially greater than the column density
of the gas emitting the far-IR CO lines (\about\ee{3}{21}\ \cmarea; Watson
\etal\ 1985) or the IR H$_2$ lines (\about\ee{3}{19}\ \cmarea; Watson
\etal\ 1985).  We derive a lower limit to the density of the
outflowing gas \nhtwo\ $>$ \ee{3}{5}\ \cmv\ from the lower
limit to the column density and a 30\as\ source diameter.  The
small size of the CO(7\to6) emission region compared with that
of the region encompassing the hot gas emitting the H$_2$ IR
rovibrational lines and the far-IR CO lines indicates that the
submillimeter CO emission lines arise in gas closer to the outflow
source and supports the contention that the gas is cooling
downstream of the hot gas shocked by the outflow.

\section {conclusions}
We observed CO(7\to6) emission across a region nearly 2~pc in
extent in the core of the Orion A cloud.  Emission is observed
out to the limits of the mapping and undoubtedly extends
further.  The line temperature of the extended CO(7\to6) emission
out to more than 8\am\ from the peak position is still about 0.2 of
the peak observed line temperature.
Except at the position of Orion~KL, the narrow
line profiles (\dv\ $<$ 7 \kms) indicate a relatively quiescent
excitation mechanism. The line intensities constrain the gas to
temperatures greater than 40~K across the entire region mapped.
Comparison of the quiescent CO(7\to6) line strengths with emission
from \cO(2\to1) indicates that the densities of the emitting gas
must be greater than $10^4$~--~$10^5$ \cmv\ and the H$_2$ column densities
must exceed $10^{21}$~--~$10^{22}$ \cmarea\ for this component.

There is good evidence that the submillimeter CO emission arises
from PDRs excited by the far-UV flux from the Trapezium cluster.
Bright 158 \um\ \cii\ line emission, which is excited in the primarily
atomic component of PDRs, radiates from the
entire region mapped in the CO(7\to6) line.  The line strength
of the quiescent CO(7\to6) emission peaks near \thetaonec\ and
indicates a gas temperature higher than 160~K, more than twice
the temperature of the dust at the same position and ruling out
collisions with dust as a heating mechanism for the gas. In
PDRs, however, UV can directly heat the gas to temperatures
exceeding the dust temperature.  The PDR models of Burton,
Hollenbach, \& Tielens (1990) can reproduce the observed
integrated intensities of the quiescent CO(7\to6) emission in the
regions a parsec or more from \thetaonec, where the UV flux is less
than a few $10^3$~\Go, as long as \nhtwo\ \gsim\ $10^5$ \cmv.

Perhaps the most interesting aspect of the CO(7\to6) spectra
is the emergence of multiple velocity components in the region
6\am\ east of \thetaonec\ and beyond.  Such a phenomenon is easily
explained if the molecular gas is highly clumped, as a number of
observations of the Orion A cloud indicate.  We interpret the
multiple velocity components as emission from two distinct
clumps or filaments separated along the line of sight, at least
one of which must be embedded in the cloud.  The presence of
bright \cii\ emission from the same region and observations of
multiple \cplus\ velocity components in carbon radio recombination
lines suggest that the CO(7\to6) lines are emitted from PDRs on
the clump or filament surfaces.  Like the \cii\ emission
from the molecular clouds associated with M17, W3, and
NGC 1977, the CO lines in Orion probably arise in PDRs throughout
the interior of the molecular cloud and not only at the surface.
Spectrally resolved observations of the 158 \um\ \cii\ line will
be needed to positively correlate the \cii\ emission with one
or both of the CO components.

Quiescent emission dominates the total flux of CO(7\to6) line
emission from Orion~A.  Although the brightest lines, with a
total velocity extent of \about100 \kms\ and brightness
temperatures exceeding 250~K, are emitted from the Orion KL
region, the compactness of the outflow source diminishes the
contribution of the broad-line emission to only a few percent of
the total flux of CO(7\to6) emission from Orion~A.  Generalizing
the results of the distribution of CO(7\to6) line flux along the
cut in R.A. through \thetaonec\ to two dimensions, we find that the
emission from regions between 6\am\ and 9\am\ from \thetaonec, where the UV
flux is less than a few $10^3$ \Go, contributes \about0.3 of the total
CO(7\to6) line flux, while emission from the brightest quiescent
lines within \about75\as\ of \thetaonec\ account for only \about0.1 of
the total line flux.  The balance of the line flux arises in the region
between 75\as\ and 6\am\ from \thetaonec.  Mapping of the CO(7\to6) line in
other molecular clouds with moderate ambient UV fields ($G$
\about100~--~1000 \Go) can determine whether widespread low-level emission
($T_B$ $<$ 50~K) dominates the total flux of CO(7\to6) emission.
A good candidate for such a mapping project is the Orion B
molecular cloud, where the ambient UV field strength varies
from \about40 to 5000 \Go\ and where bright 158 \um\ \cii\ emission
extends over most of a 24\am\by36\am\ region encompassing the H~II
region NGC 2024 and the reflection nebula NGC 2023 (Jaffe
\etal\ 1993).

The Orion A cloud offers an interesting contrast to the
more distant and more luminous star-forming regions DR~21 and
W51, where broad-line CO(7\to6) emission (\dv\ $>$ 20 \kms) is
observed to extend over 2~--~3 pc scales and dominates the total
flux emanated in the J = 7\to6 transition (Jaffe \etal\ 1989).  If
the Orion cloud were at the same distance as DR~21 or W51
(\about5 kpc), the broad-line CO(7\to6) emission from Orion~KL would
be completely unobservable due to beam dilution, given the
present sensitivity of 800 GHz receivers, and only the extended
quiescent component would be detected.

\vskip 18pt
We thank H. Dinerstein, N. Evans, D. Hollenbach, and J. Lacy for
helpful critiques of early drafts of this work.  The observations
would not have been possible without the able technical support of
the staff of the CSO.
J. E. H. acknowledges support from a NASA
Graduate Student Researchers Program fellowship.  This research
was supported by the University of Texas at Austin, NSF grant
88-17544, and a grant from the W. M. Keck Foundation.\par
{
\tenrm
\baselineskip = 12pt
\parskip = 0pt plus 0pt
\vskip 14pt
\heading {references}
\vskip 2pt
\journal {Balick, B., Gammon, R. H., \& Hjellming, R. M.} {1974} {PASP}
{86} {616}
\journal {Bally, J., Langer, W. D., Stark, A. A., \& Wilson, R. W.} {1987}
{ApJ} {312} {L45}
\journal {Batrla, W., Wilson, T. L., Bastien, P., \& Ruf, K.} {1983} {A\&A}
{128} {279}
\journal {Beckwith , S., Evans, N. J., II, Gatley, I., Gull, G., \&
Russell, R. W.} {1983} {ApJ} {264} {152}
\journal {Beckwith, S., Persson, S. E., Neugebauer, G.} {1979} {ApJ} {227}
{436}
\journal {Beckwith, S., Persson, S. E., Neugebauer, G., \& Becklin, E. E.}
{1978} {ApJ} {223} {464}
\inbook {Buhl, D., Chin, G., Koepf, G. A., Peck, D. D., \& Fetterman,
H. R.} {1982} {Submillimetre Wave Astronomy} {J. E.\ Beckman \&
J. P. Phillips} {Cambridge} {Cambridge Univ.\ Press} {111}
\journal {Blake, G. A., Sutton, E. C., Masson, C. R., \& Phillips, T. G.}
{1987} {ApJ} {315} {621}
\journal {Boreiko, R. T., Betz, A. L., \& Zmuidzinas, J.} {1988} {ApJ} {325}
{L47}
\journal {Burton, M. G., Hollenbach, D. J., \& Tielens, A. G. G. M.}
{1990} {ApJ} {365} {620}
\journal {Crawford, M. K., Genzel, R., Townes, C. H., \& Watson, D. M.}
{1985} {ApJ} {291} {755}
\journal {Erickson, N. R., Goldsmith, P. F., Snell, R. L., Berson, R. L.,
Huguenin, G. R., Ulich, B. L., \& Lada, C. J.} {1982} {ApJ} {261} {L103}
\journetal {Fetterman, H. R.} {1981} {Science} {211} {580}
\journal {Flower, D. R., \& Launay, J. M.} {1985} {MNRAS} {214} {271}
\journal {Genzel, R., Reid, M. J., Moran, J. M., \& Downes, D.} {1981} {ApJ}
{244} {884}
\journal {Genzel, R., \& Stutzki, J.} {1989} {ARAA} {27} {41}
\journal {Goldreich, P., \& Kwan, J.} {1974} {ApJ} {189} {441}
\journetal {Goldsmith, P. F.} {1981} {ApJ} {243} {L79}
\journal {Graf, U. U., Genzel, R., Harris, A. I., Hills, R. E., Russell,
A. P. G., \& Stutzki, J.} {1990} {ApJ} {358} {L49}
\journal {Harris, A. I., Stutzki, J., Genzel, R., Lugten, J. B., Stacey,
G. J., \& Jaffe, D. T.} {1987} {ApJ} {322} {L49}
\journal {Hildebrand, R. H., Loewenstein, R. F., Harper, D. A., Orton,
G.\ S., Keene, J., \& Whitcomb, S. E.} {1985} {Icarus} {64} {64}
\journal {Howe, J. E., Jaffe, D. T., Genzel, R., \& Stacey, G. J.} {1991}
{ApJ} {373} {158}
\journal {Jaffe, D. T., Davidson, J. A., Dragovan, M., \& Hildebrand, R. H.}
{1984} {ApJ} {284} {637}
\journal {Jaffe, D. T., Genzel, R., Harris, A. I., Howe, J. E., Stacey,
G. J., \& Stutzki, J.} {1990} {ApJ} {353} {193}
\journal {Jaffe, D. T., Genzel, R., Harris, A. I., Lugten, J. B., Stacey,
G. J., \& Stutzki, J.} {1989} {ApJ} {344} {265}
\journal {Jaffe, D. T., \& Pankonin, V.} {1978} {ApJ} {226} {869}
\privcom {Jaffe, D. T., \etal} {1993} {in preparation}
\journal {Koepf, G. A., Buhl, D., Chin, G., Peck, D. D., Fetterman, H. R.,
Clifton, B. J., \& Tannenwald, P. E.} {1982} {ApJ} {260} {584}
\journal {Kr\"autle, H., Sauter, E., \& Schultz, G. V.} {1977} {IR Phys.} {17}
{477}
\journal {Kutner, M. L., \& Ulich, B. L.} {1981} {ApJ} {250} {341}
\journal {Lellouch, E., Encrenaz, T., \& Combes, M.} {1984} {A\&A} {140} {405}
\journetal {Matsuhara, H.} {1989} {ApJ} {339} {L67}
\journal {McKee, C. F., Storey, J. W. V., Watson, D. M., \& Green, S.}
{1982} {ApJ} {259} {647}
\journal {Mundy, L. G., Cornwell, T. J., Masson, C. R., Scoville, N. Z.,
Baath, L. B., \& Johansson, L. E. B.} {1988} {ApJ} {325} {382}
\journal {Mundy, L. G., Scoville, N. Z., Baath, L. B., Masson, C. R., \&
Woody, D. P.} {1986} {ApJ} {304} {L51}
\journal {Nadeau, D., \& Geballe, T. R.} {1979} {ApJ} {230} {L169}
\journal {Penzias, A. A., \& Burrus, C. A.} {1973} {ARAA} {11} {51}
\journal {Schinke, R., Engel, V., Buch, U., Meyer, H., \& Diercksen,
G. H. F.} {1985} {ApJ} {299} {939}
\journetal {Schmid-Burgk, J.} {1989} {A\&A} {215} {150}
\journal {Schultz, G. V., Durwen, E. J., R\"oser, H. P., Sherwood, W. A.,
\& Wattenbach, R.} {1985} {ApJ} {291} {L59}
\journal {Scoville, N. Z., Hall, D. N. B., Kleinmann, S. G.,
\& Ridgway, S. T.} {1982} {ApJ} {253} {136}
\journal {Scoville, N. Z., \& Solomon, P. M.} {1974} {ApJ} {187} {L67}
\journal {Stacey, G. J., Viscuso, P. J., Fuller, C. E., \& Kurtz, N. T.}
{1985} {ApJ} {289} {803}
\journal {Stacey, G. J., Jaffe, D. T., Geis, N., Genzel, R., Harris, A. I.,
Poglitsch, A., Stutzki, J., \& Townes, C. H.} {1993} {ApJ} {404} {000}
\journal {Sternberg, A., \& Dalgarno, A.} {1989} {ApJ} {338} {197}
\journal {Stutzki, J., Stacey, G. J., Genzel, R., Harris, A. I., Jaffe,
D. T., \& Lugten, J. B.} {1988} {ApJ} {332} {379}
\journal {Tauber, J. A., \& Goldsmith, P. F.} {1990} {ApJ} {356} {L63}
\journal {Tielens, A. G. G. M., \& Hollenbach, D.} {1985a} {ApJ} {291} {722}
\journal {Tielens, A. G. G. M., \& Hollenbach, D.} {1985b} {ApJ} {291} {747}
\journal {van Dishoeck, E. F., \& Black, J. H.} {1988} {ApJ} {334} {771}
\journal {Watson, D. M., Genzel, R., Townes, C. H., \& Storey, J. W. V.}
{1985} {ApJ} {298} {316}
\journal {Werner, M. W., Crawford, M. K., Genzel, R., Hollenbach, D. J.,
Townes, C. H., \& Watson, D. M.} {1984} {ApJ} {282} {L81}
\journal {Werner, M. W., Gatley, I., Harper, D. A., Becklin, E. E.,
Loewenstein, R. F., Telesco, C. M., \& Thronson, H. A.} {1976}
{ApJ} {204} {420}
\journal {Wolfire, M. G., Hollenbach, D., \& Tielens, A. G. G. M.} {1989}
{ApJ} {344} {770}
\journal {Wright, E. L.} {1976} {ApJ} {210} {250}
\journal {Wright, E. L., \& Odenwald, S.} {1980} {BAAS} {12} {456}
\journal {Ziurys, L. M., Martin, R. N., Pauls, T. A., \& Wilson, T. L.}
{1981} {A\&A} {104} {288}
\journal {Zuckerman, B.} {1973} {ApJ} {183} {163}
}
\bye